\documentclass[runningheads]{llncs}

\usepackage{graphicx}
\usepackage{hyperref}
\usepackage{booktabs}
\usepackage{amsmath}
\usepackage{amssymb}
\usepackage[dvipsnames]{xcolor}
\usepackage[framemethod=TikZ]{mdframed}

\begin{document}

\titlerunning{Toward Self-Driving Universities}

\title{Toward Self-Driving Universities:\\
\large Can Universities Drive Themselves with Agentic AI?}

\author{Anis Koubaa}

\institute{
AlfaisaX: Center of Cognitive Robotics and Autonomous Agents, Alfaisal University, Riyadh, Saudi Arabia \\
\email{akoubaa@alfaisal.edu}
}

\maketitle

\begin{abstract}
The rapid evolution of Agentic AI and large language models (LLMs) presents transformative opportunities for higher education institutions. This chapter introduces the concept of \textit{self-driving universities}---a vision where AI-enabled systems progressively automate administrative, academic, and quality-assurance processes following a progressive autonomy model inspired by self-driving systems. We examine the current challenges facing traditional universities, including bureaucratic overload, fragmented information systems, and the disproportionate time faculty spend on clerical tasks---diverting effort away from timely feedback, curricular improvement, student mentorship, and research productivity. While prior AI-in-education research has focused primarily on learning support, tutoring, and analytics, there is a lack of system-level frameworks for automating institutional quality assurance and accreditation workflows using agentic AI. We address this gap by presenting a framework for progressive automation, detailing how agentic AI can transform course design, assessment alignment, accreditation documentation, and institutional reporting. Through case studies of pilot deployments, we demonstrate that AI-assisted workflows can substantially reduce task completion times while enabling capabilities previously considered infeasible. The chapter's originality lies in introducing an autonomy-level framework for higher education operations, grounded in agentic AI architectures rather than prompt-based LLM assistance. Finally, we discuss the critical infrastructure requirements, ethical considerations, and a roadmap for universities to transition toward higher levels of academic autonomy.

\keywords{Self-Driving Universities \and Agentic AI \and Large Language Models \and Academic Quality Assurance \and Higher Education Automation \and Accreditation}
\end{abstract}


\section{Introduction: The Case for Self-Driving Universities}

\subsection{The Crisis of Administrative Overload in Higher Education}

Contemporary higher education institutions operate within an increasingly complex ecosystem of regulatory compliance, quality assurance frameworks, and bureaucratic oversight. This administrative expansion has fundamentally altered the nature of academic work. Longitudinal studies indicate that faculty members now allocate a substantial majority of their professional time to non-instructional activities, including documentation, reporting, and compliance verification~\cite{Woelert2023AdminBurden}. This structural shift represents a significant departure from the traditional academic mission centered on teaching, research, and student engagement.

The administrative burden manifests most acutely in the context of institutional accreditation. Bodies such as the Accreditation Board for Engineering and Technology (ABET) and the National Center for Academic Accreditation and Evaluation (NCAAA) require systematic evidence of learning outcome achievement, program effectiveness, and continuous improvement~\cite{ABET2025EACCriteria}. While these frameworks serve legitimate quality assurance purposes, their implementation generates substantial operational overhead. The preparation of self-study reports, course specification documents, and comprehensive assessment records can consume weeks or months of faculty effort per accreditation cycle. This creates a tension between compliance obligations and the academic activities these frameworks are designed to enhance.

The proliferation of fragmented information systems compounds these challenges. Academic data frequently resides in disconnected repositories—learning management systems, student information systems, departmental spreadsheets—preventing unified institutional oversight~\cite{warren2025enhancing}. This fragmentation necessitates extensive manual data aggregation and reconciliation, introducing opportunities for error and version inconsistency. The resulting inefficiency imposes measurable opportunity costs: diminished research productivity, reduced instructional innovation, and elevated professional burnout among faculty and administrative staff.

\subsection{The Self-Driving Analogy: Autonomy as a Spectrum}

The concept of the \textit{self-driving university} draws a deliberate analogy to autonomous vehicle technology, which has adopted a well-established taxonomy of automation levels~\cite{Khan2022Level5Survey}. In automotive contexts, levels range from Level 0 (no automation) through Level 5 (full autonomy in all conditions). This framework proves instructive for conceptualizing academic automation not as a binary transformation but as a progressive transition across a spectrum of institutional capabilities.

At Level 0, all academic and administrative processes rely entirely on human execution. Level 1 introduces basic assistive technologies—template-based document generation, spell-checking utilities—that augment but do not replace human judgment. Level 2 encompasses partial automation, where artificial intelligence systems generate preliminary content or recommendations subject to human review and modification. Level 3 represents conditional automation, wherein AI agents manage routine workflows while humans retain supervisory authority for exception handling. Level 4 denotes high automation, where AI systems independently execute complex processes with minimal human intervention, and Level 5 envisions fully autonomous academic operations with self-optimizing institutional systems.

This graduated framework serves two critical functions. First, it provides a structured roadmap for institutional transformation, enabling universities to pursue automation incrementally rather than through disruptive wholesale replacement. Second, it acknowledges that different academic functions may warrant different levels of autonomy. Routine compliance documentation may appropriately operate at Level 4, while decisions involving academic judgment or student welfare may require sustained human oversight characteristic of Level 2 or 3.

The analogy also foregrounds a crucial insight: autonomy requires robust infrastructure. Autonomous vehicles depend on high-definition mapping, sensor fusion, and real-time processing capabilities. Similarly, self-driving universities require unified data architectures, interoperable systems, and validated knowledge bases~\cite{Gao2023RAGSurvey}. The prerequisite for effective AI deployment is not merely algorithmic sophistication but systematic institutional preparedness.

\subsection{Chapter Objectives and Scope}

This chapter examines the technical foundations, practical implementations, and governance implications of progressive academic automation through agentic AI systems. The term ``agentic AI'' refers to artificial intelligence architectures capable of autonomous goal pursuit through iterative planning, tool use, memory persistence, and self-evaluation~\cite{acharya2025agentic,kostopoulos2025agentic}. Unlike passive generative models that respond to isolated prompts, agentic systems decompose complex institutional objectives into executable workflows, adapt responses based on contextual feedback, and maintain coherence across extended task sequences.

The chapter pursues three primary objectives. First, it establishes a conceptual framework for self-driving universities, mapping AI capabilities onto the autonomy levels previously introduced and identifying target domains for automation. Second, it presents evidence from operational implementations, including systems for automated assessment~\cite{Impey_2024,XEducation2026}, accreditation documentation~\cite{Muhamad2025NCAAA}, and peer review management~\cite{ReviewPilot2024}. These case studies provide qualitative insights into implementation benefits and challenges. Third, it addresses governance requirements, including data infrastructure prerequisites, ethical considerations surrounding algorithmic decision-making, and change management strategies for organizational adoption.

The scope is deliberately focused on administrative and quality assurance processes rather than pedagogical interventions. While AI applications in personalized learning and intelligent tutoring warrant attention, this chapter concentrates on institutional operations—accreditation workflows, assessment documentation, compliance verification—where automation can demonstrably reduce faculty burden without compromising academic judgment. The analysis draws primarily on implementations within engineering and computer science programs, where structured learning outcomes and quantifiable assessment criteria facilitate systematic evaluation.

This chapter does not advocate for the elimination of human roles in academic governance. Rather, it proposes a cognitive offloading model wherein routine, data-intensive tasks are delegated to AI systems, enabling faculty and administrators to focus on strategic decision-making, student mentorship, and scholarly inquiry. The self-driving university, properly conceived, represents an augmentation of institutional capacity rather than its replacement.

\section{Challenges in Traditional University Operations}

Before examining how agentic AI can transform academic administration, it is essential to understand the structural inefficiencies that characterize contemporary university operations. This section analyzes four interconnected challenges: bureaucratic complexity, accreditation burden, information fragmentation, and the consequent erosion of faculty productivity.

\subsection{Bureaucratic Complexity in Academic Institutions}

Modern universities operate through layered administrative hierarchies that have expanded substantially over recent decades. Course approval processes exemplify this complexity: a single curriculum modification may require sequential review by department committees, college councils, and institutional governance bodies, with each stage introducing documentation requirements and approval delays. What might constitute a straightforward pedagogical improvement---updating a course to reflect current industry practices---can require months of administrative processing before implementation.

This bureaucratic expansion reflects legitimate institutional needs: accountability, quality oversight, and regulatory compliance. However, the cumulative effect creates operational friction that impedes academic responsiveness. Redundant documentation requirements compound the challenge. Faculty members preparing course materials must often populate multiple systems with overlapping information: learning management platforms, course specification templates, departmental records, and accreditation databases. Each system maintains its own format requirements, creating duplicative effort without commensurate benefit.

Siloed departmental operations further exacerbate these inefficiencies. Academic units frequently develop localized procedures, documentation standards, and approval workflows that diverge from institutional norms. While such autonomy supports disciplinary specificity, it fragments institutional knowledge and complicates cross-departmental coordination. A faculty member transferring between departments, or a program seeking to establish interdisciplinary courses, encounters procedural inconsistencies that multiply administrative burden.

\subsection{The Quality Assurance and Accreditation Burden}

Institutional accreditation represents one of the most resource-intensive administrative functions in contemporary higher education. Accrediting bodies such as ABET for engineering programs and regional agencies like NCAAA impose comprehensive requirements for demonstrating educational quality~\cite{ABET2025EACCriteria}. While these frameworks serve essential purposes---ensuring program rigor, protecting student interests, and maintaining institutional credibility---their implementation generates substantial operational demands.

Course specification compliance illustrates the scale of documentation required. Each course must articulate learning outcomes aligned with program-level objectives, assessment strategies mapped to specific competencies, and continuous improvement mechanisms informed by systematic data collection. For a typical engineering program with thirty to forty courses, maintaining current and compliant specifications represents a considerable documentation effort. When accreditation cycles require comprehensive self-study reports---documents that may span hundreds of pages with supporting evidence appendices---the resource commitment intensifies further.

The mapping of Course Learning Outcomes (CLOs) to Program Learning Outcomes (PLOs) constitutes a particularly demanding requirement. Accrediting bodies expect institutions to demonstrate not merely that courses address program objectives, but that assessment instruments validly measure student achievement of these outcomes. This necessitates systematic tracking of student performance across multiple assessment points, aggregation of results to program-level indicators, and analysis of trends to inform curricular decisions~\cite{Muhamad2025NCAAA}. Manual execution of these processes is both time-consuming and error-prone, particularly when data resides across disconnected systems.

The Quality Assurance Agency (QAA) and similar bodies have recognized these challenges, noting that institutional capacity for quality assurance is increasingly strained by expanding requirements~\cite{qaa2024quality}. The Tertiary Education Quality and Standards Agency (TEQSA) has similarly emphasized the need for assessment reform to address the growing complexity of compliance obligations~\cite{teqsa2024risk}.

\subsection{Fragmented Information Systems}

A fundamental obstacle to efficient university operations---and a critical barrier to effective AI deployment---is the fragmentation of institutional information systems. Academic data typically resides across multiple platforms: student information systems manage enrollment and grades, learning management systems host course content and assessments, research administration systems track grants and publications, and departmental databases maintain program-specific records. These systems frequently operate as isolated silos with limited interoperability.

The consequences of this fragmentation are multifaceted. First, data redundancy introduces inconsistency risks. When the same information---course descriptions, faculty credentials, student records---must be entered into multiple systems, version conflicts become inevitable. A curriculum change updated in one system but not propagated to others creates discrepancies that complicate reporting and decision-making. Second, analytical capabilities are constrained. Without unified data architecture, institutions lack the integrated visibility necessary for evidence-based strategic planning. Answering seemingly straightforward questions---Which courses contribute most effectively to program outcomes? Where do students struggle across the curriculum?---requires manual data aggregation from disparate sources.

For AI applications, information fragmentation presents a particularly acute challenge. Machine learning systems and large language models depend on comprehensive, consistent datasets for effective operation. An AI system designed to automate accreditation documentation cannot function optimally if required data is scattered across incompatible platforms with inconsistent formatting. The principle of ``garbage in, garbage out'' applies with particular force: AI systems trained on or querying incomplete, inconsistent, or outdated data will produce correspondingly unreliable outputs~\cite{warren2025enhancing}. Consequently, any serious effort toward academic automation must address underlying data infrastructure as a prerequisite, not an afterthought.

\subsection{The Hidden Cost: Faculty Time and Productivity}

The aggregate effect of bureaucratic complexity, accreditation burden, and information fragmentation manifests most visibly in faculty time allocation. Research on administrative burden in higher education indicates that faculty members dedicate substantial portions of their professional effort to non-instructional activities~\cite{Woelert2023AdminBurden}. Documentation, reporting, compliance verification, and committee service consume time that might otherwise support teaching innovation, research productivity, or student mentorship.

This reallocation of faculty effort carries significant opportunity costs. Research output---publications, grant acquisition, scholarly impact---depends on sustained intellectual engagement that administrative tasks interrupt. Teaching quality suffers when course preparation competes with documentation requirements for limited faculty attention. Student advising and mentorship, activities central to educational mission, are curtailed when faculty schedules are dominated by compliance obligations.

The professional consequences extend beyond productivity metrics. Chronic administrative burden contributes to faculty burnout, reduced job satisfaction, and increased turnover intention. When skilled academics perceive that bureaucratic demands prevent them from executing core professional responsibilities, institutional culture and morale deteriorate. The irony is evident: quality assurance mechanisms designed to enhance educational outcomes may inadvertently undermine them by diverting faculty attention from the educational activities those mechanisms are intended to improve.

These observations do not constitute an argument against accountability or quality assurance. Rather, they highlight the need for more efficient mechanisms---systems that can satisfy legitimate compliance requirements while preserving faculty capacity for academic work. It is precisely this need that agentic AI systems are positioned to address, as subsequent sections will demonstrate.

\section{Agentic AI and LLMs as Enablers of University Automation}

The administrative challenges outlined in the previous section have persisted despite decades of digitization efforts. Traditional software systems---rule-based workflows, template engines, and database queries---proved insufficient because academic processes involve natural language interpretation, contextual judgment, and policy reasoning that resist rigid automation. Recent advances in artificial intelligence, particularly large language models and agentic architectures, offer fundamentally different capabilities. This section examines three enabling technologies: transformer-based language models, retrieval-augmented generation, and agentic AI systems.

\subsection{The Transformer Revolution: Technical Foundations}

The transformer architecture, introduced in 2017, represents a paradigm shift in how machines process language. Unlike previous approaches that analyzed text sequentially, transformers employ a mechanism called \textit{self-attention} that allows simultaneous consideration of all words in a document and their relationships to one another~\cite{Yao2022ReAct}.

\begin{figure}[h!]
\centering
\includegraphics[width=0.75\textwidth]{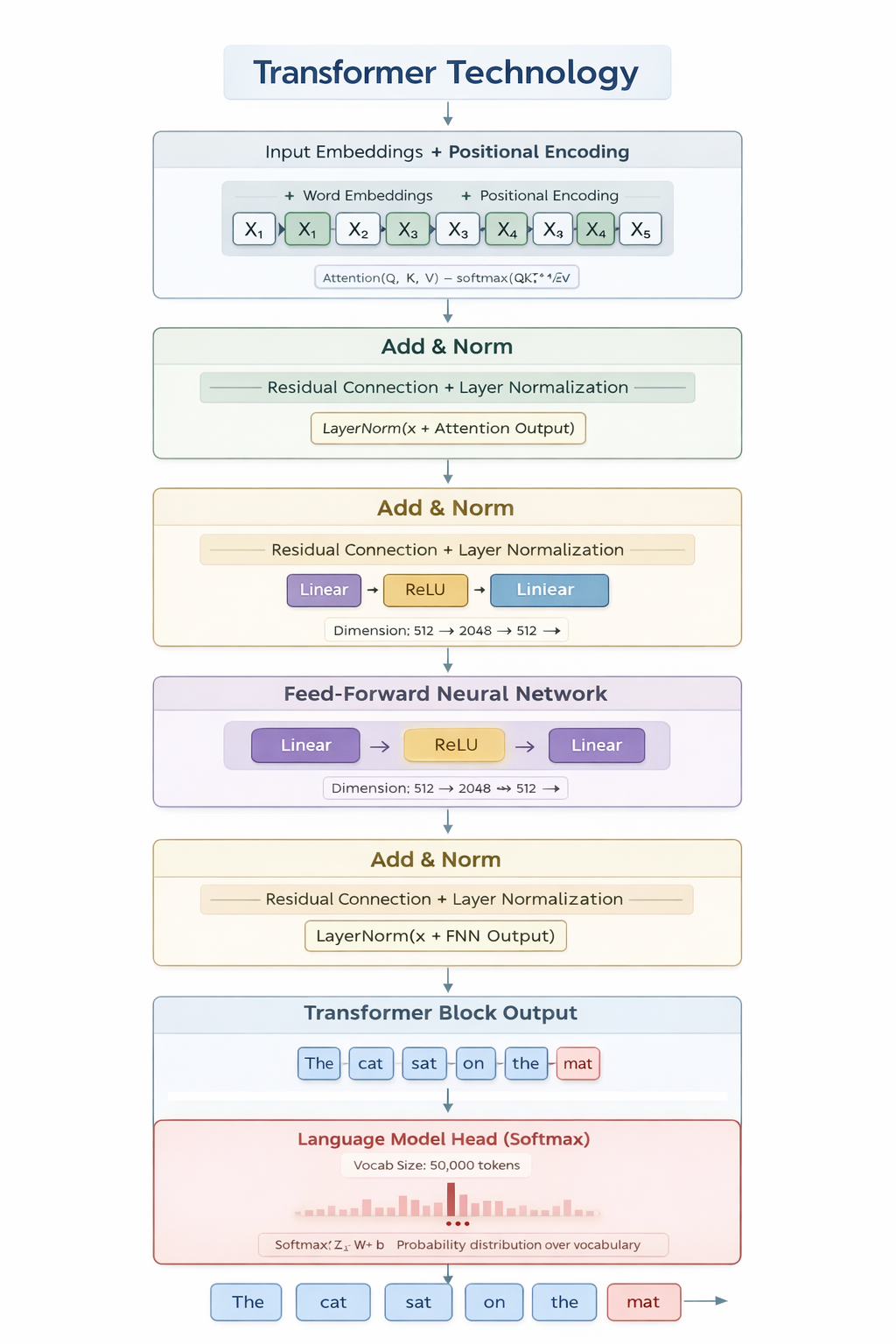}
\caption{Self-attention mechanism in transformer architecture. Each input token attends to all other tokens, enabling contextual understanding of relationships within academic text.}
\label{fig:transformer}
\end{figure}

Figure~\ref{fig:transformer} illustrates this mechanism. When processing a course description such as ``Students will analyze algorithms and design efficient solutions,'' the self-attention mechanism computes relevance scores between every word pair. The model learns that ``analyze'' relates strongly to ``algorithms'' and that ``design'' connects to ``solutions''---contextual relationships that inform appropriate responses.

This architecture enables \textit{probabilistic intelligence} rather than deterministic rule-following. When a faculty member asks an AI system to ``align this exam question with ABET Student Outcome 2,'' the model draws on patterns learned from millions of examples rather than executing predefined rules. This flexibility proves essential for academic contexts where policies contain ambiguity, exceptions require interpretation, and appropriate responses depend on disciplinary conventions that resist exhaustive codification.

\subsection{Retrieval-Augmented Generation for Institutional Knowledge}

While transformer models possess broad linguistic capabilities, they lack access to institution-specific information: university bylaws, accreditation manuals, course catalogs, and departmental policies. Retrieval-Augmented Generation (RAG) addresses this limitation by grounding model responses in verified institutional documents~\cite{Gao2023RAGSurvey}.

\begin{figure}[h!]
\centering
\includegraphics[width=0.85\textwidth]{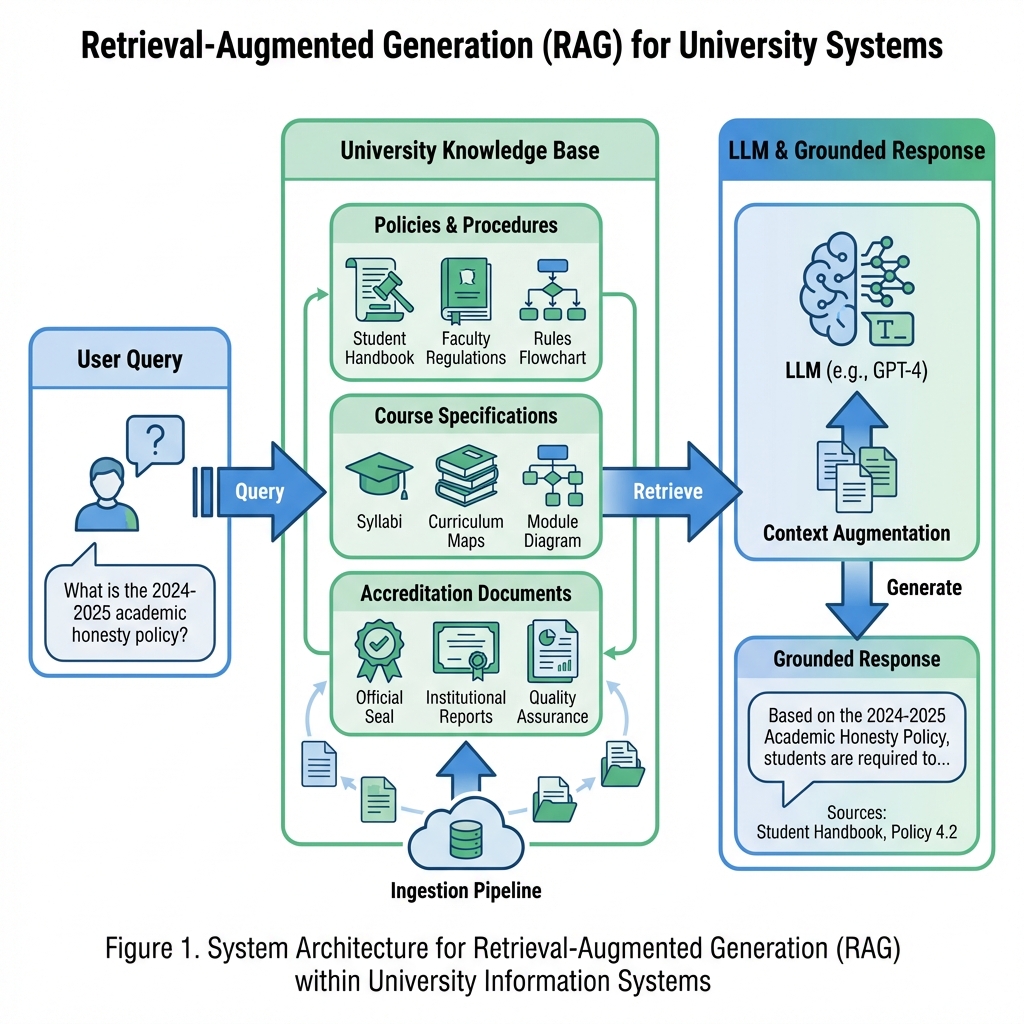}
\caption{Retrieval-Augmented Generation architecture for university systems. User queries trigger document retrieval from the institutional knowledge base, grounding LLM responses in verified policies and specifications.}
\label{fig:rag}
\end{figure}

Figure~\ref{fig:rag} depicts this architecture. When a user queries the system---for example, ``What are the prerequisites for CS301?''---the RAG pipeline first searches a knowledge base containing university documents. Relevant passages are retrieved and provided to the language model alongside the original query. The model then generates a response that synthesizes retrieved information, citing specific policy sources rather than relying solely on parametric knowledge.

This approach directly addresses the hallucination problem---the tendency of language models to generate plausible but factually incorrect information. By constraining responses to retrieved content, RAG systems can cite specific university policies, quote accreditation criteria, and provide traceable justifications. For compliance-sensitive academic processes, this grounding is essential: an AI system drafting accreditation documentation must cite actual institutional evidence, not invent supporting claims.

\subsection{From Generative AI to Agentic AI}

Standard generative AI operates reactively: a user provides a prompt, the model generates a response, and the interaction concludes. This pattern, while useful for drafting and brainstorming, proves insufficient for complex administrative workflows that require multiple steps, verification, and iterative refinement. \textit{Agentic AI} extends generative capabilities with autonomous goal pursuit~\cite{acharya2025agentic,kostopoulos2025agentic}.

\begin{figure}[h!]
\centering
\includegraphics[width=0.75\textwidth]{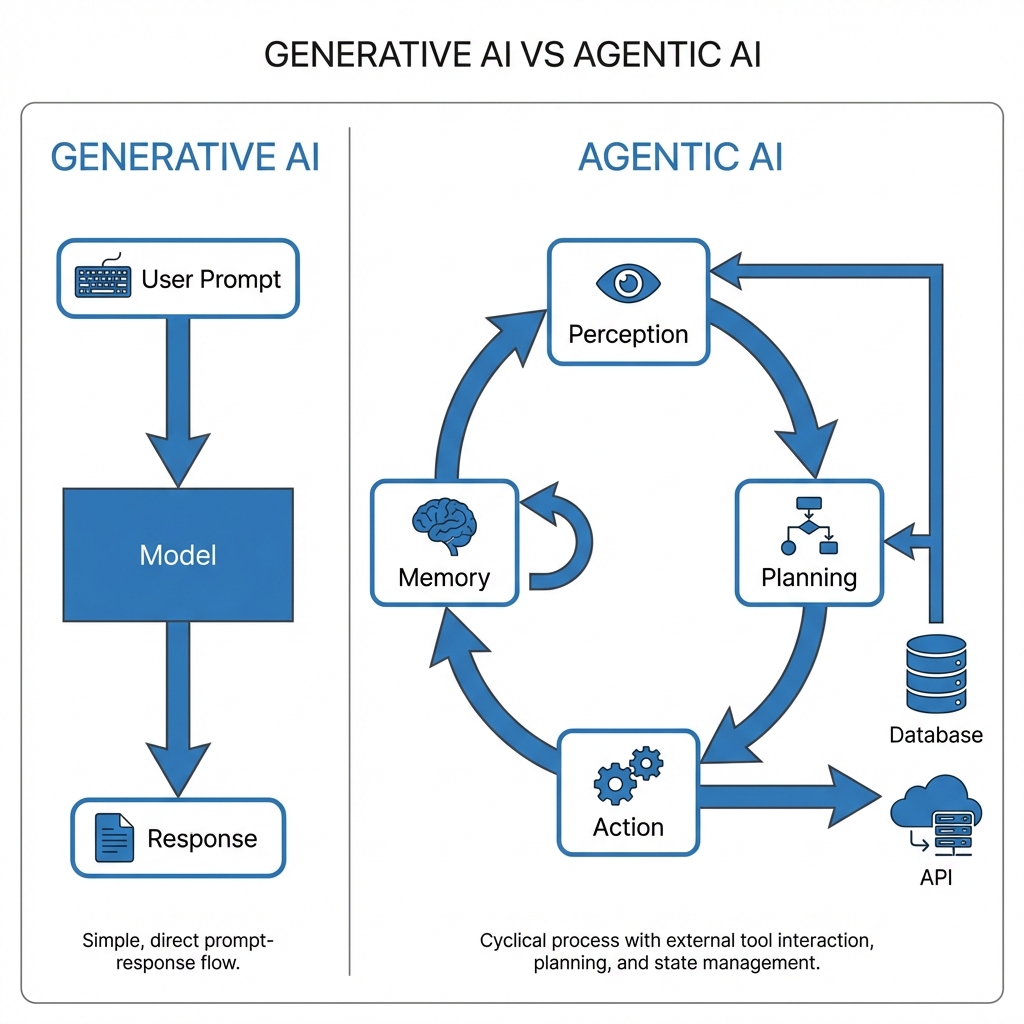}
\caption{Generative AI versus Agentic AI. While generative systems follow simple prompt-response patterns, agentic systems execute iterative perception-planning-action-memory loops with external tool integration.}
\label{fig:agentic}
\end{figure}

Figure~\ref{fig:agentic} contrasts these paradigms. Generative AI follows a linear prompt-response pattern. Agentic AI, by contrast, operates through iterative loops incorporating four cognitive components. \textit{Perception} interprets the current state and task requirements. \textit{Planning} decomposes complex goals into executable sub-tasks. \textit{Action} executes steps, including invocation of external tools such as databases and APIs. \textit{Memory} maintains context across extended task sequences, enabling coherent multi-step workflows.

Consider an accreditation documentation task: ``Prepare the course folder for CS301 including outcomes assessment and improvement actions.'' A generative system might draft a single document. An agentic system would decompose this goal: retrieve course specifications from the curriculum database, extract grade distributions from the learning management system, calculate CLO achievement percentages, identify outcomes below threshold, query previous improvement plans for context, draft recommended actions, and compile the complete folder---verifying each component before proceeding.

The ReAct framework (Reason + Act) formalizes this behavior through alternating reasoning and action steps~\cite{Yao2022ReAct}. Tools extend agent capabilities: Toolformer-style models can invoke calculators, search engines, and institutional APIs~\cite{Schick2023Toolformer}. Crucially, these architectures support human-in-the-loop validation, where agents pause for approval before consequential actions---precisely the oversight mechanism required for Level 3 and Level 4 autonomy in the self-driving university framework.

\section{A Framework for Self-Driving Universities}

The preceding sections established both the problem---administrative burden consuming faculty capacity---and the solution components---transformer-based language models, retrieval-augmented generation, and agentic AI architectures. This section synthesizes these elements into a coherent framework for progressive academic automation, defining autonomy levels, identifying target domains, and proposing a transition roadmap.

\subsection{Autonomy Levels in Higher Education}

\begin{figure}[h!]
\centering
\includegraphics[width=0.85\textwidth]{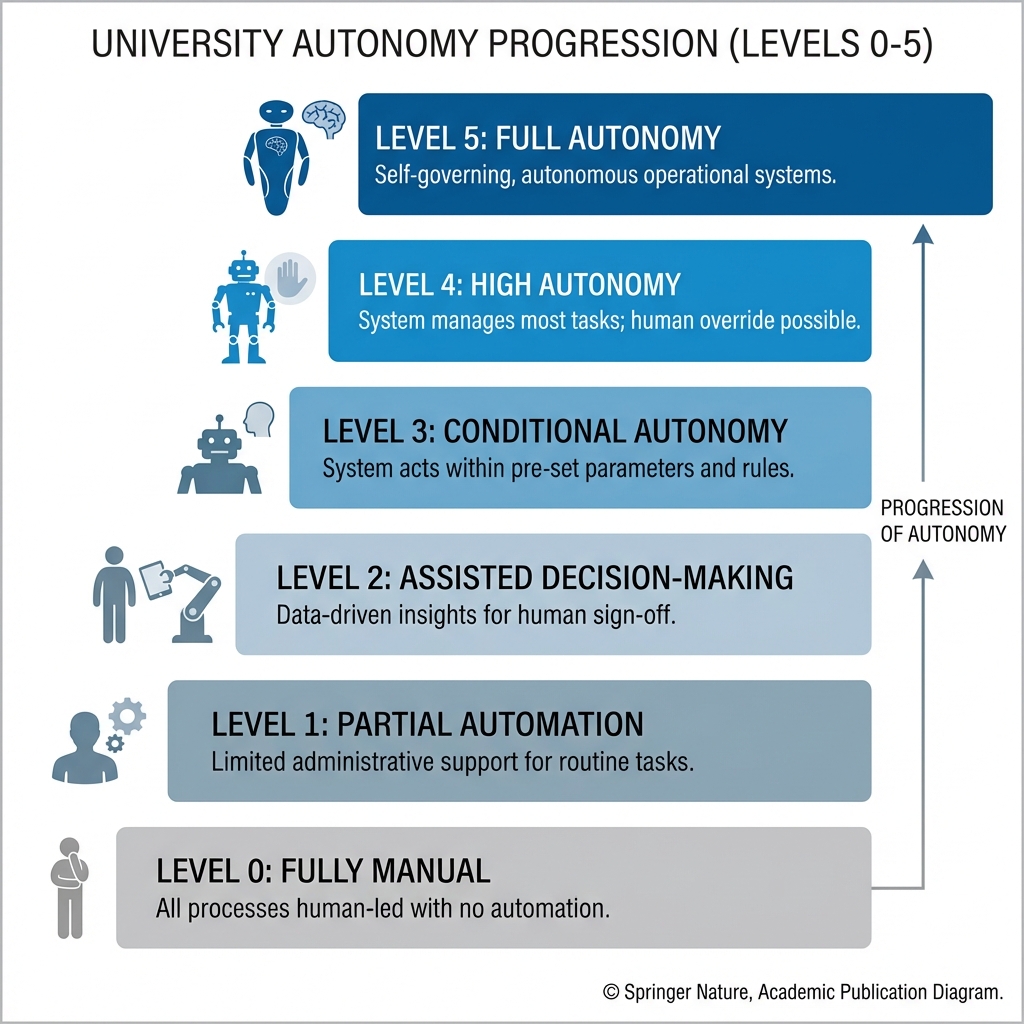}
\caption{University autonomy progression from Level 0 (fully manual) to Level 5 (full autonomy). Each level represents increasing AI capability and decreasing human intervention requirements.}
\label{fig:autonomy}
\end{figure}

Drawing on the established taxonomy from autonomous vehicle development~\cite{Khan2022Level5Survey}, we propose six levels of academic autonomy ranging from fully manual operations to self-governing institutional systems. Figure~\ref{fig:autonomy} illustrates this progression.

\textbf{Level 0: Fully Manual.} All academic and administrative processes rely entirely on human execution. Faculty manually prepare course materials, grade assessments, document outcomes, and compile compliance reports without technological assistance. This represents the historical baseline against which automation benefits are measured.

\textbf{Level 1: Basic Assistance.} Simple tools augment human effort without replacing judgment. Spell-checkers, document templates, grade calculators, and learning management system features reduce mechanical effort while humans retain complete control over content and decisions. Most contemporary institutions operate at this level for the majority of their processes.

\textbf{Level 2: Partial Automation.} AI systems generate preliminary outputs subject to human review and modification. A faculty member might request draft learning outcomes, receive AI-generated suggestions, and refine them based on disciplinary expertise. The system assists but does not decide; humans evaluate, accept, reject, or modify all AI contributions.

\textbf{Level 3: Conditional Automation.} AI agents manage routine workflows autonomously within defined parameters, with humans retaining supervisory authority for exceptions and quality assurance. An agent might automatically populate course folders with assessment data, calculate CLO achievement percentages, and flag outcomes below threshold---pausing for faculty approval before finalizing recommendations. This level represents the practical near-term target for most academic processes.

\textbf{Level 4: High Automation.} AI systems independently execute complex multi-step processes with minimal human intervention, though humans retain override capability and periodic validation responsibilities. An accreditation documentation system at this level might compile complete self-study reports from institutional data, verify compliance with each criterion, and produce submission-ready documents that humans review for accuracy rather than construct from scratch.

\textbf{Level 5: Full Autonomy.} Self-governing academic systems that learn, adapt, and optimize without routine human involvement. At this speculative level, institutional AI would not merely execute prescribed processes but would identify improvement opportunities, implement curricular adjustments based on outcome data, and maintain accreditation compliance proactively. This level remains aspirational and raises substantial governance questions that Section 6 addresses.

Table~\ref{tab:autonomy} summarizes these levels with representative academic applications.

\begin{table}[h!]
\centering
\caption{Autonomy levels in academic operations with representative applications.}
\label{tab:autonomy}
\begin{tabular}{p{1.5cm}p{3cm}p{4cm}p{3.5cm}}
\toprule
\textbf{Level} & \textbf{Description} & \textbf{Academic Example} & \textbf{Human Role} \\
\midrule
0 & Fully Manual & Hand-grading all exams & Execute all tasks \\
1 & Basic Assistance & Using rubric templates & Execute with tool support \\
2 & Partial Automation & AI drafts course descriptions & Review and refine \\
3 & Conditional & AI compiles course folders & Supervise and approve \\
4 & High Automation & AI generates self-study reports & Validate and override \\
5 & Full Autonomy & Self-optimizing curriculum & Strategic oversight only \\
\bottomrule
\end{tabular}
\end{table}

\subsection{Target Domains for Automation}

Not all academic functions are equally amenable to automation. We identify five primary domains where agentic AI can deliver substantial efficiency gains while respecting the boundaries of appropriate human oversight.

\textbf{Academic Processes.} Course design, curriculum development, and student advising involve structured knowledge that AI systems can leverage effectively. Agents can generate course specifications aligned with program outcomes, recommend prerequisite sequences based on content analysis, and provide personalized advising based on student academic records and career objectives.

\textbf{Assessment Processes.} Exam generation, grading, and outcome mapping represent high-volume, structured tasks where AI assistance proves particularly valuable~\cite{Impey_2024}. Systems can generate assessment items aligned with specific learning outcomes and Bloom's taxonomy levels, grade constructed responses against defined rubrics, and map student performance to CLO achievement metrics.

\textbf{Quality Assurance.} Compliance verification, CLO-PLO alignment checking, and continuous improvement documentation constitute prime automation targets~\cite{Muhamad2025NCAAA}. AI agents can verify that course specifications satisfy accreditation criteria, identify gaps in outcome coverage across curricula, and generate improvement action recommendations based on assessment data trends.

\textbf{Reporting.} Course folders, accreditation documentation, and institutional reports require systematic data aggregation from multiple sources---precisely the capability that agentic systems provide. Agents can compile required evidence, format outputs according to agency specifications, and maintain version-controlled documentation repositories.

\textbf{Research Support.} Literature review, proposal drafting, and peer review management benefit from AI augmentation~\cite{ReviewPilot2024}. Systems can synthesize relevant prior work, identify methodological precedents, and manage reviewer assignment and deadline tracking for scholarly publication processes.

\subsection{The Data Infrastructure Prerequisite}

The analogy to autonomous vehicles illuminates a critical constraint: even the most advanced self-driving systems have not achieved full Level 5 autonomy in all conditions. Similarly, universities cannot simply deploy AI agents and expect autonomous operation. The enabling prerequisite is comprehensive, consistent data infrastructure.

The challenge in most institutions is fragmented information systems. Student attendance resides in one database, learning management in another, faculty portfolios in a third, and registration records in a fourth. When different departments are asked the same question---How many students are enrolled this semester?---they may provide conflicting answers because each system maintains its own version of institutional truth. This inconsistency fundamentally undermines AI effectiveness: agents cannot reason correctly over contradictory data.

The first step toward a self-driving university is therefore not deploying AI agents but building the unified data foundation they require. A pilot implementation at a partner institution illustrates this approach: centralizing data for all stakeholders (students, faculty, administrators), organizational structures, facilities (buildings, classrooms, laboratories), academic programs, and individual courses. This centralized repository enables consistent study plan management, term scheduling, enrollment tracking, and attendance monitoring. Only with this foundation in place can AI agents reliably automate advising, course compliance checking, and report generation.

This observation reinforces a key message: data infrastructure investment is not a technical convenience but a strategic prerequisite. Institutions that attempt to deploy agentic AI atop fragmented systems will encounter the ``garbage in, garbage out'' problem identified in Section 2.3, undermining both efficiency gains and stakeholder trust in AI-assisted processes.

\subsection{The Transition Roadmap}

Institutional transformation toward self-driving operations proceeds through four phases. Critically, the first phase focuses entirely on data infrastructure---not AI deployment. Attempting to implement agentic AI atop fragmented, incomplete, or inconsistent information systems yields partial efficiency at best and ``garbage in, garbage out'' failures at worst. The roadmap therefore begins with foundation-building before any automation.

\textbf{Phase 1: Infrastructure Foundation.} The essential first phase establishes a complete and consistent information ecosystem before introducing AI capabilities. This phase addresses the fragmentation problem directly:
\begin{itemize}
\item \textit{Data Inventory and Audit}: Catalog all institutional data sources---student information systems, learning management platforms, faculty records, facilities databases, registration systems---and document inconsistencies.
\item \textit{Unified Data Architecture}: Design and implement a centralized repository or data lake that consolidates information across departments, establishing a single source of truth.
\item \textit{Data Quality Assurance}: Resolve conflicts (e.g., different departments reporting different enrollment figures), standardize formats, and implement validation rules.
\item \textit{API Development}: Create programmatic interfaces that enable future AI systems to access institutional data securely and consistently.
\end{itemize}

This foundational philosophy drives the development of platforms like XEducation (discussed in Section 5), which was designed explicitly to create a unified ecosystem. By integrating critical entities---institutions, buildings, classrooms, programs, courses, departments, colleges, employees, students, and academic terms---into a single coherent data model, XEducation eliminates the fragmentation that plagues traditional systems. This consistency provides the solid infrastructure required for agentic AI to operate effectively, ensuring that automated decisions are based on complete and accurate institutional reality.

Only when the institution possesses a comprehensive, consistent view of its own ecosystem---students, faculty, courses, programs, facilities, outcomes---does AI deployment become viable. Expected duration: 12--24 months.

\textbf{Phase 2: Tool Augmentation.} With data infrastructure in place, the second phase introduces AI as an optional assistant available to faculty and administrators. Users experiment with generative capabilities for drafting course materials, brainstorming assessment items, and querying institutional data through natural language interfaces. This phase builds familiarity, identifies high-value use cases, and surfaces usability concerns that inform subsequent planning. AI operates as a tool that users invoke deliberately; no processes are restructured. Expected duration: 12--18 months.

\textbf{Phase 3: Process Automation.} The third phase integrates AI agents into institutional workflows as co-pilots. Standard processes incorporate automated steps while retaining human checkpoints: agents draft course folders that faculty review, calculate CLO achievement metrics that coordinators validate, and generate compliance reports that administrators verify before submission. Training programs develop faculty competencies in effective AI collaboration. Governance frameworks establish oversight protocols, quality benchmarks, and escalation procedures for AI-assisted decisions. Expected duration: 18--36 months.

\textbf{Phase 4: Autonomous Operations.} The mature phase transitions AI from co-pilot to manager for appropriate processes, with humans assuming supervisory rather than executive roles. Routine documentation, compliance verification, and reporting operate autonomously within defined parameters. Human expertise concentrates on strategic decisions, exception handling, and activities requiring interpersonal judgment---teaching, mentorship, and research that constitute the core academic mission. This phase represents ongoing evolution rather than a terminal state, with continuous refinement of autonomy boundaries based on performance data and stakeholder feedback.

\section{Case Studies: Agentic AI in Academic Practice}

The theoretical framework described in the previous section is currently being operationalized through pilot deployments at partner institutions. This section presents evidence from an initial deployment and reports qualitative observations from the implementation.

\subsection{Pilot Deployment: MUST University, Tunisia}

The primary case study derives from a pilot deployment at MUST University (Mediterranean University of Science and Technology), a private institution in Tunisia operating under both U.S.-style academic standards and Tunisian Ministry of Higher Education requirements.

\subsubsection{Context and Institutional Adoption}

MUST University adopted XEducation, a unified information system developed in partnership with the university. The institutional decision was motivated by the challenges described in Section 2: inconsistent data across departments, manual compilation of accreditation documentation, and faculty time consumed by administrative tasks rather than teaching and research.

The deployment integrated all institutional data into a single coherent ecosystem: academic terms, programs, courses, syllabi, faculty records, student enrollment, attendance tracking, examination records, and grade management. This unified foundation---consistent with the Phase 1 infrastructure requirements outlined in Section 4.4---established the prerequisite for subsequent AI agent deployment.

\subsubsection{AI-Assisted Data Ingestion}

A significant early efficiency gain came from AI-assisted data entry. Rather than manually completing form fields for course specifications and syllabi, faculty upload existing documents (PDFs, Word files). Large language models extract structured information---learning outcomes, topic sequences, assessment methods, textbook references---and populate the database automatically. Human review confirms accuracy before finalization.

This approach substantially reduced the time required to enter a complete course specification, transforming a task that previously consumed a significant portion of faculty time into a brief document upload and verification process. The reduction is particularly significant during curriculum updates, where dozens of specifications require simultaneous modification.

\subsubsection{Dual Grading System Reconciliation}

MUST University operates under a unique constraint: it uses U.S.-style letter grading (A, A+, B, etc.) for internal academic purposes while simultaneously complying with Tunisian Ministry requirements, which mandate numeric grading scales. Previously, this dual system required manual transcription and conversion, creating both workload and error risk.

XEducation automates this reconciliation. A single grade entry triggers automatic conversion to both scales, with configurable mapping rules. Transcripts and ministry reports are generated in parallel from the same underlying data, eliminating discrepancies and reducing administrative overhead.

\subsection{Agentic AI Applications in the Pilot}

With the unified data infrastructure in place, several agentic AI capabilities were deployed within the pilot environment.

\subsubsection{ExamGPT: Assessment Creation, Grading, and CLO Analysis}

One of the most labor-intensive aspects of accreditation compliance is the calculation and documentation of Course Learning Outcome (CLO) achievement. Traditional approaches require faculty to manually aggregate assessment data, calculate achievement percentages, and compile reports showing student performance against each outcome. This process is repeated for every course, every semester, consuming substantial faculty time and introducing opportunities for calculation errors.

The ExamGPT module---a core component of the XEducation platform---addresses this burden while operating with full context awareness of the institutional course catalog. When a faculty member requests exam generation, the agent accesses the specific course's learning outcomes, topic coverage, and Bloom's taxonomy alignment directly from the integrated database. This contextual grounding produces more relevant assessment items than generic LLM prompting.

\begin{figure}[!ht]
\centering
\includegraphics[width=0.95\textwidth]{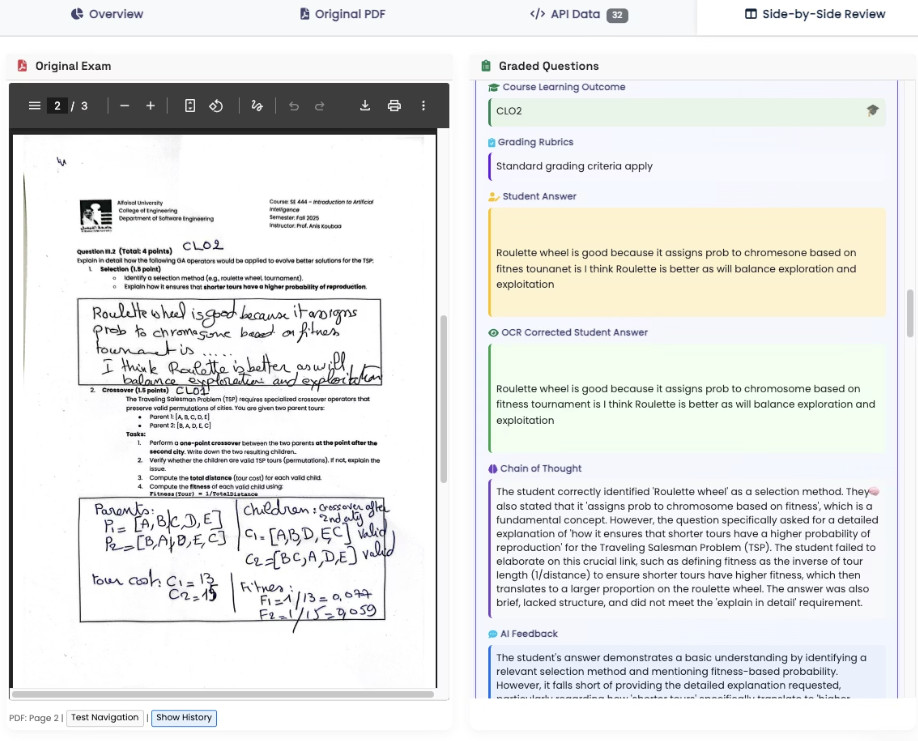}
\caption{Vision-language model processing of handwritten examination. The interface shows the original scanned document alongside OCR-corrected text, CLO mapping, chain-of-thought reasoning, and AI-generated feedback.}
\label{fig:handwritten-grading}
\end{figure}

ExamGPT supports two distinct grading modalities: traditional handwritten examinations and online form-based assessments. For handwritten exams, vision-language models process scanned examination papers, performing optical character recognition (OCR) on student handwriting and evaluating responses against rubrics. Figure~\ref{fig:handwritten-grading} illustrates this side-by-side review interface: the left panel displays the original scanned document with handwritten answers, while the right panel shows the OCR-corrected text, the mapped Course Learning Outcome, the AI's chain-of-thought reasoning, and generated feedback. This transparency in AI reasoning enables instructors to understand and verify the basis for suggested grades.

\begin{figure}[!ht]
\centering
\includegraphics[width=0.95\textwidth]{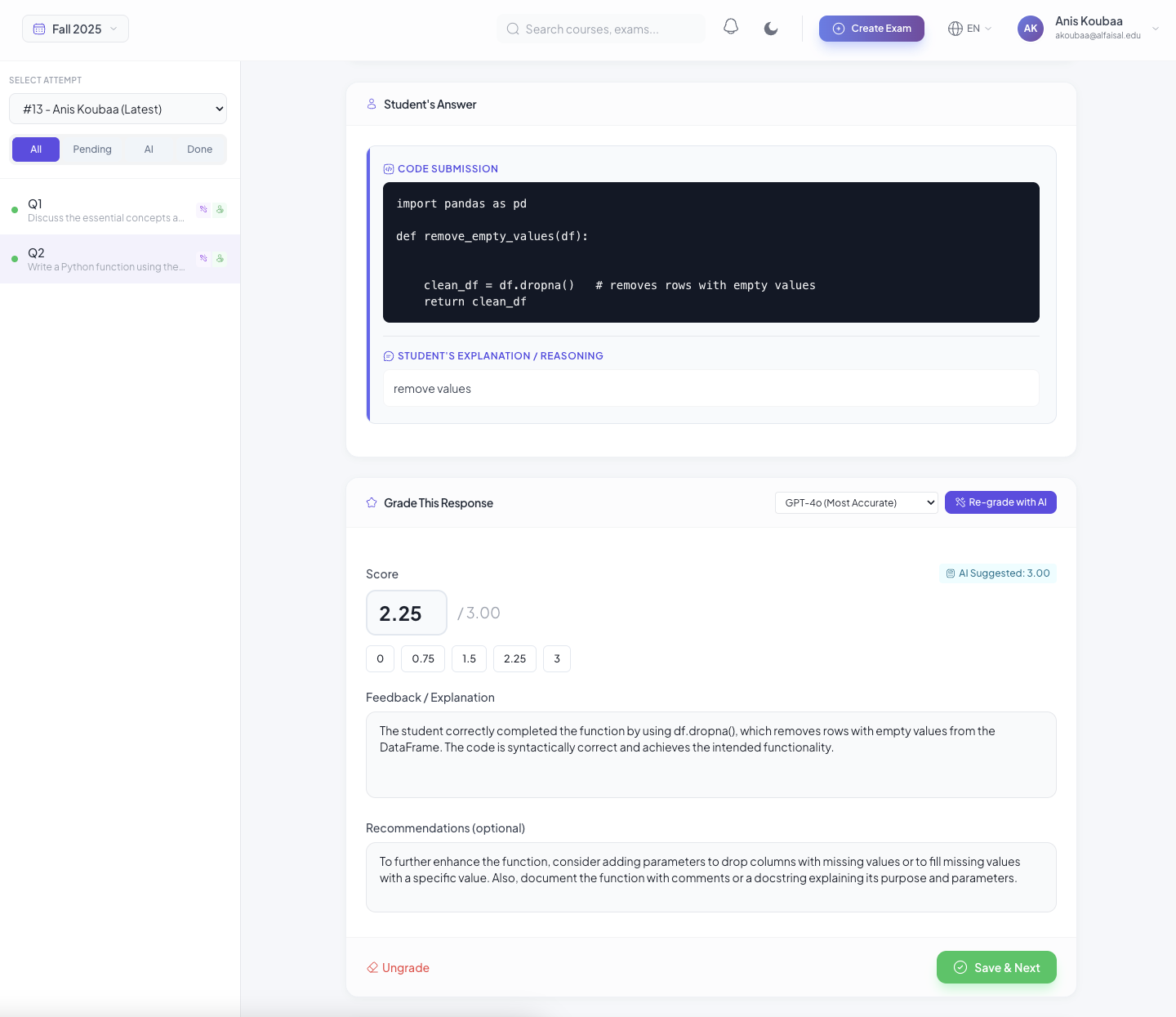}
\caption{AI-assisted grading of an online coding question. The system suggests a score and generates feedback, with the instructor retaining authority to approve or modify the assessment.}
\label{fig:grading-coding}
\end{figure}

\begin{figure}[!ht]
\centering
\includegraphics[width=0.95\textwidth]{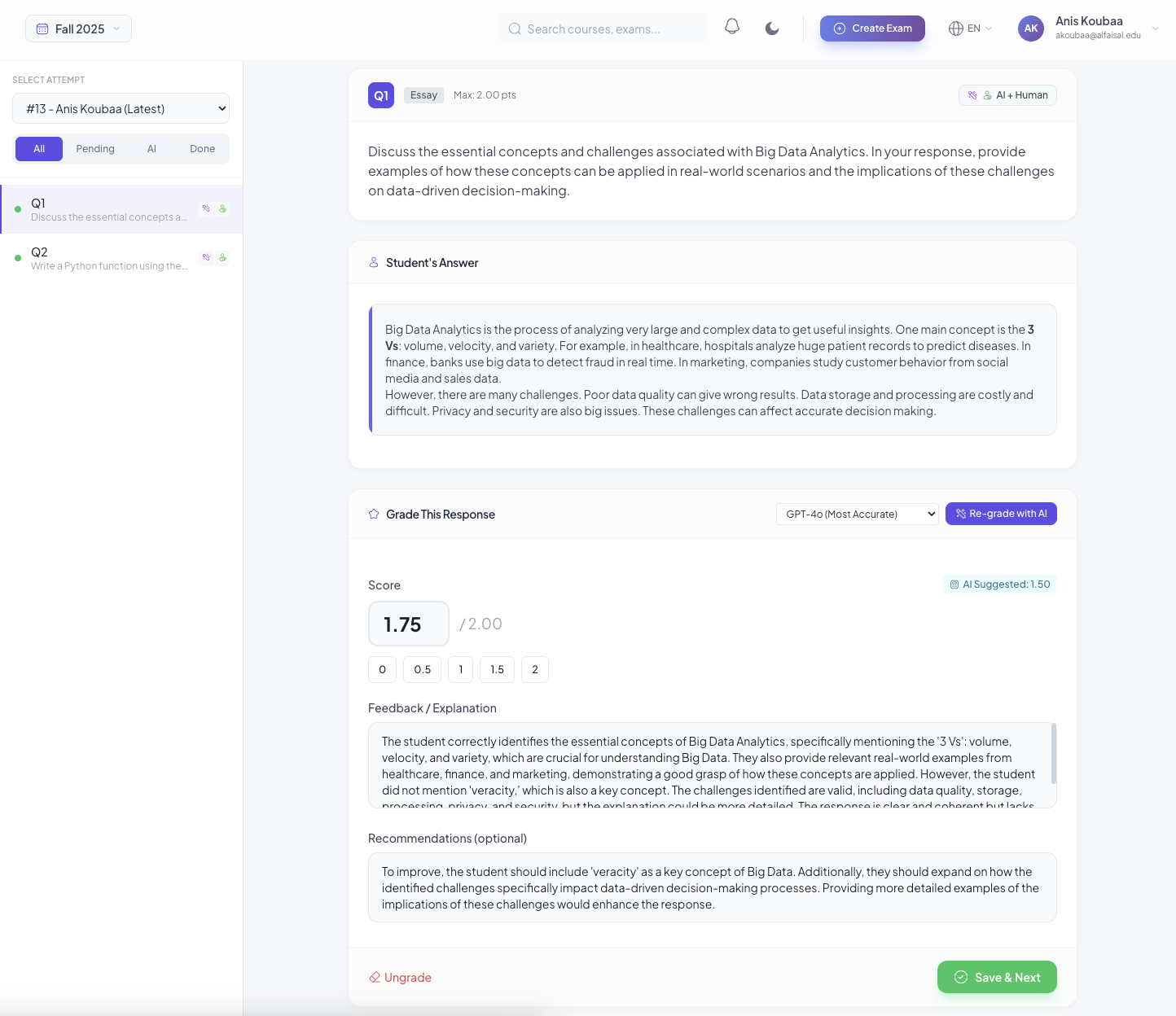}
\caption{AI-assisted grading of an online essay question. The ``AI + Human'' indicator ensures assessment provenance is transparent, showing both AI-suggested and human-approved scores.}
\label{fig:grading-essay}
\end{figure}

For online assessments, ExamGPT applies the same analytical capabilities to form-based submissions including programming exercises and essay questions. Figure~\ref{fig:grading-coding} illustrates grading of a coding question: the AI analyzes the student's code submission, evaluates correctness and style, and suggests a score with detailed feedback. Figure~\ref{fig:grading-essay} demonstrates essay grading, where the AI provides content analysis, identifies strengths and areas for improvement, and offers recommendations for the student.

Across both modalities, ExamGPT implements a \textit{human-in-the-loop} architecture rather than fully autonomous grading. Recognizing the ethical imperative of human oversight in high-stakes assessment, the instructor retains full authority: they can accept the AI suggestion, adjust the score, modify the feedback, or override the assessment entirely. The ``AI + Human'' badge visible in the interface indicates that this assessment combines machine intelligence with human judgment.

This hybrid approach addresses several governance concerns. First, ExamGPT maintains full traceability: for each graded question, the system records whether the assessment was performed by AI alone, human alone, or collaboratively. This audit trail is essential for accreditation compliance and academic integrity investigations. Second, the architecture ensures that AI serves as an accelerator rather than a replacement---reducing grading time while preserving faculty authority over academic judgment. Third, the explicit visibility of AI involvement addresses transparency requirements increasingly mandated by institutional policies and quality assurance frameworks.

Once grades are finalized, each score is linked to specific CLOs, enabling automatic calculation of CLO achievement statistics. This transforms grading from a purely administrative task into an immediate source of outcome assessment data.

\begin{figure}[!ht]
\centering
\includegraphics[width=0.95\textwidth]{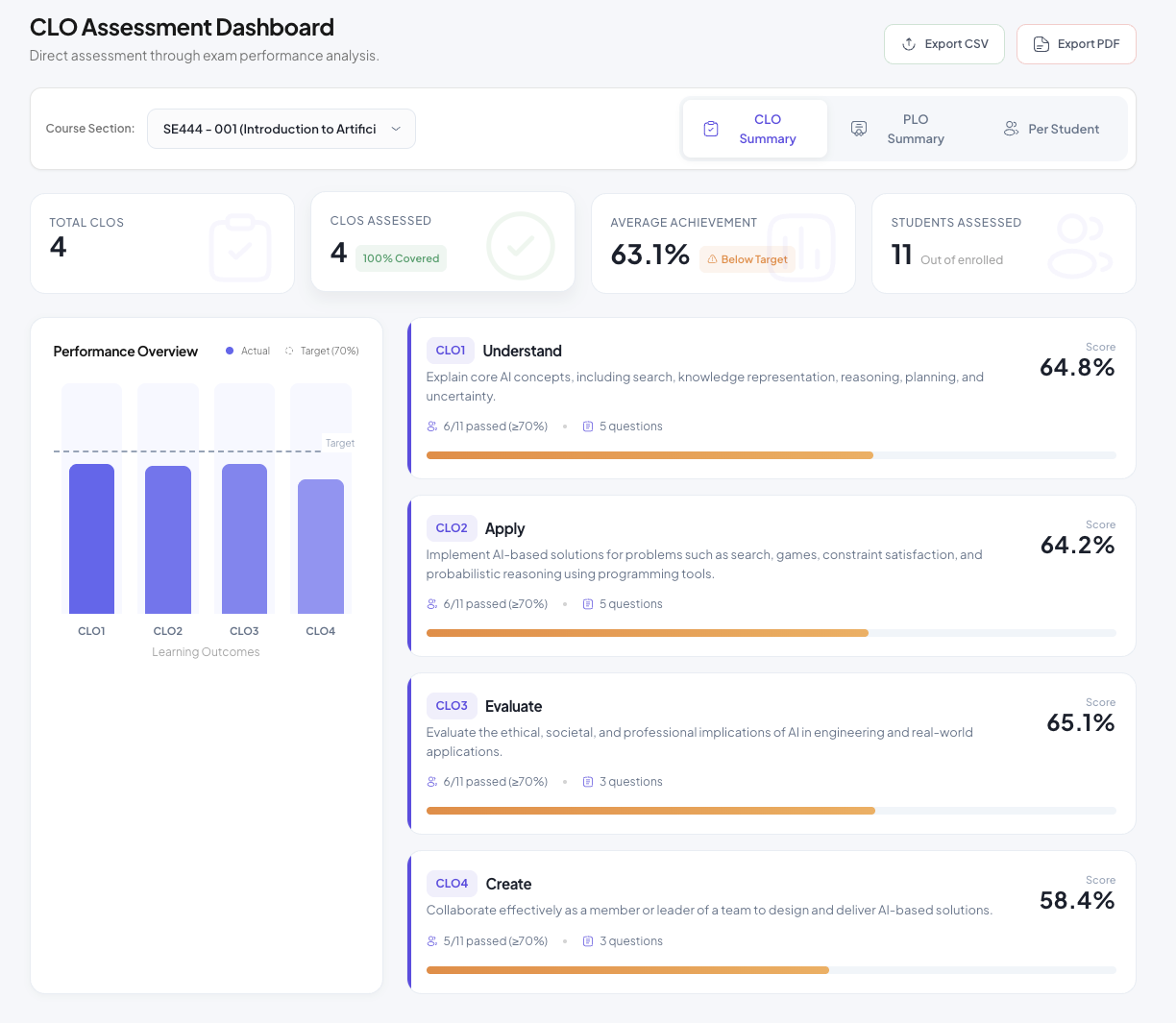}
\caption{CLO Assessment Dashboard showing per-outcome achievement percentages, Bloom's taxonomy alignment, and performance visualization. Data is aggregated automatically from ExamGPT grading.}
\label{fig:clo-dashboard}
\end{figure}

Because all grades flow into the unified XEducation database with CLO linkages, the system automatically generates real-time analytics. Figure~\ref{fig:clo-dashboard} illustrates the CLO Assessment Dashboard, which displays achievement statistics for each learning outcome. The system calculates the percentage of students meeting the target threshold for each CLO, identifies outcomes requiring attention, and provides drill-down capability to individual assessment items. What traditionally required faculty to manually aggregate data from spreadsheets now updates automatically with each graded assessment.

\subsubsection{Program-Level Outcome Aggregation}

Beyond course-level CLO assessment, the ExamGPT module within XEducation extends its analytical capabilities to program-level outcome aggregation. One of the most labor-intensive aspects of accreditation compliance is mapping course outcomes to program outcomes and calculating aggregate PLO achievement. Traditional approaches require program coordinators to manually compile matrices from individual course reports---a process consuming substantial time and introducing calculation errors.

\begin{figure}[!ht]
\centering
\includegraphics[width=0.95\textwidth]{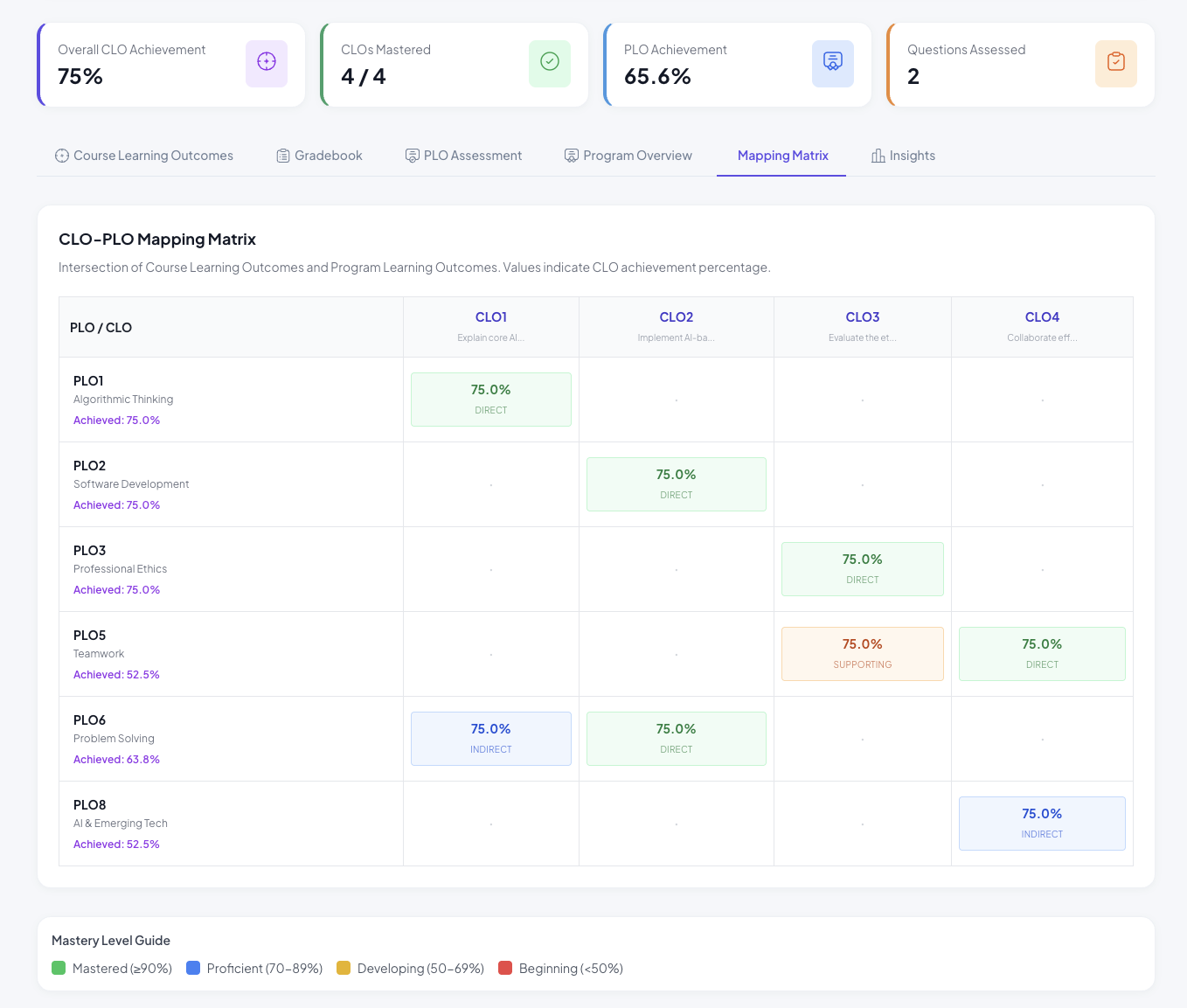}
\caption{CLO--PLO Mapping Matrix displaying the intersection of course learning outcomes with program learning outcomes. Achievement percentages and contribution types are calculated automatically by the ExamGPT module within XEducation.}
\label{fig:clo-plo-matrix}
\end{figure}

With unified infrastructure, these calculations become fully automated. The CLO--PLO mapping matrix, shown in Figure~\ref{fig:clo-plo-matrix}, demonstrates how course-level achievements propagate to program-level outcomes. Each cell indicates the contribution type (direct, indirect, or supporting) and the achievement percentage. This matrix---which traditionally requires hours of manual compilation---is generated automatically from the underlying assessment data.

\begin{figure}[!ht]
\centering
\includegraphics[width=0.95\textwidth]{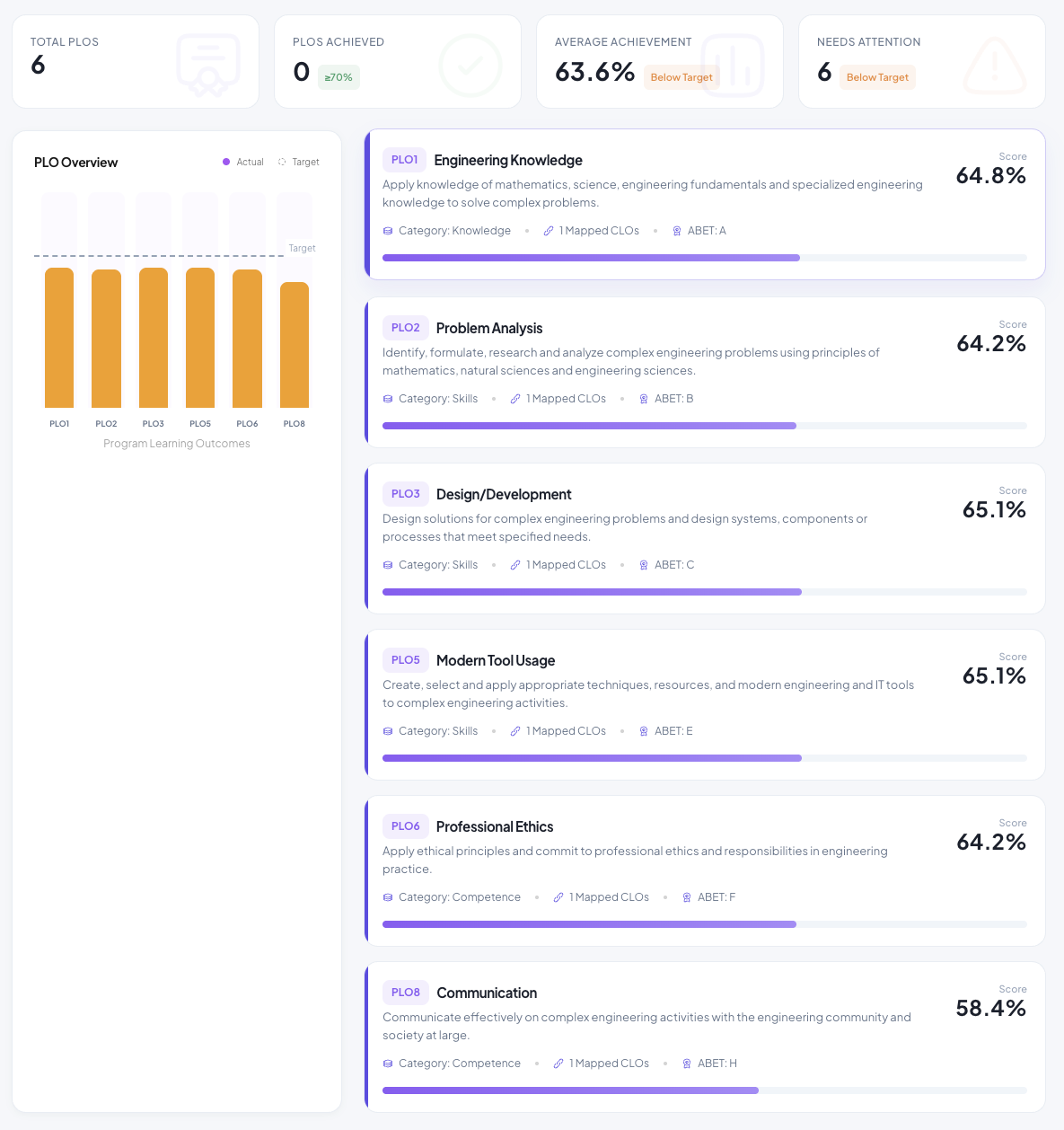}
\caption{PLO Assessment Dashboard showing program-level outcome achievement with ABET criterion alignment. The ExamGPT module automatically aggregates data from all contributing courses within the XEducation platform.}
\label{fig:plo-dashboard}
\end{figure}

At the program level, the PLO Assessment Dashboard (Figure~\ref{fig:plo-dashboard}) aggregates data across all courses to provide a comprehensive view of program outcome achievement. Each PLO is linked to its corresponding ABET criterion, enabling direct verification of accreditation compliance. The system flags outcomes below threshold, enabling targeted intervention before accreditation visits.

The automation of CLO and PLO assessment represents a paradigm shift in quality assurance workload. What previously consumed dozens of faculty hours per course per semester---manual data entry, spreadsheet calculations, matrix compilation, and report formatting---now occurs continuously and automatically. Faculty review dashboards rather than construct them, focusing their expertise on interpreting results and designing interventions rather than on data aggregation.

\subsubsection{Course Quality Assurance and Accreditation Review}

AI agents review course specifications against accreditation standards (ABET, NCAAA, or institutional criteria). The system identifies missing components, flags misaligned outcomes, and suggests remediation. Previously, such compliance checking required manual review by quality assurance staff---a bottleneck during accreditation preparation cycles.

In the pilot, course specification review time decreased dramatically from manual review requiring substantial faculty attention to rapid AI-assisted verification with human confirmation. Given that a typical program contains dozens of courses, the aggregate time savings during accreditation preparation is substantial.

\subsubsection{Strategic Analytics via Agentic Reasoning}

Beyond document-level automation, the pilot deployed agentic workflows for institutional analytics. University administrators can pose natural language queries---``Show me attendance patterns across all first-year courses, highlighting courses below target attendance thresholds''---and receive structured reports.

The underlying agentic workflow decomposes the query into data retrieval steps, executes queries against the unified database, aggregates results across organizational units, and formats the output for human consumption. What previously required a data analyst to compile manually is now available rapidly through automated workflows. This capability transforms raw institutional data into actionable intelligence for strategic decision-making.

\subsubsection{Automated Insight Reports via Agentic Workflows}

A distinguishing capability of the XEducation platform is its integration of agentic AI workers that autonomously generate institutional insight reports. These workers operate in two complementary modes: \textit{predefined workflows} that execute on scheduled triggers or event-based conditions, and \textit{on-demand analysis} powered by ReAct (Reasoning and Acting) agents that interpret natural language requests, plan multi-step analyses, and synthesize findings into structured reports.

Three categories of automated reports demonstrate this capability. The \textbf{Daily Health Check} workflow runs automatically each evening, aggregating attendance data across all courses and departments. If institutional attendance falls below target thresholds, the system drills down into problem areas, identifies at-risk courses, and generates recommendations for department heads. The \textbf{At-Risk Student Detection} workflow can be triggered weekly (batch scan) or in real-time when a student accumulates consecutive absences beyond acceptable limits. It analyzes the student's attendance profile, identifies behavioral patterns, performs cohort comparison, and recommends appropriate interventions based on risk level (critical, warning, watch, or safe). The \textbf{Comparative Analysis} workflow enables administrators to compare any two groups---morning versus afternoon classes, Department A versus Department B---providing comparative analysis and actionable interpretation.

The following box illustrates a sample AI-generated report for at-risk student detection:

\begin{mdframed}[
  linecolor=Purple,
  linewidth=2pt,
  roundcorner=8pt,
  backgroundcolor=Purple!5,
  innertopmargin=10pt,
  innerbottommargin=10pt
]
\textbf{\textcolor{Purple}{At-Risk Student Report}} \\[2pt]
\textit{Generated: 2025-12-13 05:23:02 | Type: student\_insight} \\[8pt]

\textbf{\textcolor{ForestGreen}{About This Report}} \\
This report is generated by the \textbf{At-Risk Student Detection} workflow. It can be triggered weekly (batch scan of all students) or in real-time when a student misses 3 consecutive sessions. The workflow analyzes the student's attendance profile, course breakdown, behavioral patterns (by day/hour), and cohort comparison to determine risk level and recommend appropriate interventions. \\[8pt]

\begin{tabular}{@{}p{4cm}p{6cm}@{}}
\textbf{Risk Assessment} & \\
\quad Risk Level: & \textcolor{Orange}{\textbf{WARNING}} \\
\quad Cohort Percentile: & Bottom 30\% \\[6pt]
\textbf{Attendance Data} & \\
\quad Overall Rate: & 72.8\% \\
\quad Total Absences: & 61 \\
\quad Problem Courses: & Introduction to Digital Systems, Academic English, Introduction to Computer Systems, Mathematics I \\[6pt]
\textbf{Behavioral Pattern} & \\
\quad Peak Absences: & Monday mornings, Friday afternoons \\
\quad Trend: & Declining over past 4 weeks \\[6pt]
\textbf{Recommended Actions} & \\
\multicolumn{2}{@{}p{10cm}@{}}{1. Schedule meeting with academic advisor within 48 hours. 2. Notify course instructors of risk status. 3. Consider referral to student support services.} \\
\end{tabular}
\end{mdframed}

These reports exemplify the paradigm shift from reactive to proactive institutional management. Rather than waiting for end-of-semester grade reports to identify struggling students, the system continuously monitors behavioral indicators and surfaces concerns in time for meaningful intervention. The agentic architecture ensures that reports are not merely data aggregations but include contextual interpretation, pattern analysis, and actionable recommendations---the kind of synthesis that previously required experienced analysts to produce.

\subsection{Qualitative Observations}

The pilot deployment at MUST University provided qualitative evidence of the framework's practical value. Faculty and administrators reported substantial reductions in time spent on routine administrative tasks, allowing them to redirect effort toward teaching, mentorship, and research activities.

\textbf{Observed Benefits.} Several categories of improvement emerged from the pilot:
\begin{itemize}
\item \textit{Course specification management}: Faculty reported that AI-assisted data ingestion transformed course specification updates from time-consuming manual processes into brief verification tasks.
\item \textit{Learning outcome alignment}: Tasks such as drafting CLOs and mapping them to Bloom's taxonomy levels---previously requiring careful manual construction---were accelerated substantially through AI-generated drafts subject to human refinement.
\item \textit{Assessment workflows}: Exam creation, rubric development, and grading assistance reduced faculty workload, particularly for courses with large enrollments.
\item \textit{Compliance preparation}: Accreditation documentation that previously consumed significant faculty time during review cycles was compiled more rapidly through automated aggregation of course folder contents.
\item \textit{Institutional analytics}: Cross-program reports that previously required manual data compilation were generated on-demand through natural language queries.
\end{itemize}

\textbf{Limitations.} These observations derive from a single institutional pilot with favorable preconditions: unified data infrastructure, institutional commitment to adoption, and technical support during deployment. Generalization to other contexts requires attention to the Phase 1 infrastructure prerequisites discussed in Section 4.4. The benefits observed here demonstrate feasibility rather than guaranteed outcomes across all institutional contexts.

\section{Challenges, Ethical Considerations, and the Road Ahead}

The preceding sections have presented agentic AI as a promising avenue for addressing the administrative burden in higher education. However, responsible adoption requires acknowledging significant challenges that remain unresolved. This section examines infrastructure prerequisites, ethical considerations, organizational resistance, and directions for future research.

\subsection{Infrastructure and Data Governance}

As emphasized throughout this chapter, agentic AI cannot function effectively without a unified data foundation. Institutions with fragmented systems---separate databases for registration, learning management, and quality assurance---will realize only partial benefits from AI deployment. Data quality issues compound this problem: inconsistent records, missing fields, and outdated information lead to unreliable agent outputs.

Beyond technical integration, institutions must establish clear data governance policies. Questions of data ownership, access permissions, and retention periods require explicit resolution before AI agents access sensitive student and faculty records. Interoperability standards remain underdeveloped in higher education, making cross-system integration more costly than it should be.

\subsection{Ethical and Governance Considerations}

Academic integrity concerns arise when AI assists in assessment. If students can use generative AI to complete assignments, the validity of traditional evaluation methods diminishes~\cite{teqsa2024risk}. Similarly, if AI grades student work, questions of fairness and appeal mechanisms require careful design.

Transparency presents another challenge. When an AI agent recommends that a student take a particular course or flags a faculty member's CLO achievement as below threshold, stakeholders require explanations. Current LLMs lack robust interpretability, making it difficult to justify automated decisions to affected parties.

Human oversight remains essential. The framework presented in this chapter deliberately limits full autonomy (Level 5) to aspirational status, recognizing that consequential academic decisions---degree conferral, faculty evaluation, accreditation compliance---require human accountability. Bias in training data may propagate into automated recommendations, potentially disadvantaging certain student populations. Institutions must implement monitoring mechanisms to detect and correct such disparities.

\subsection{Resistance and Change Management}

Technological capability alone does not ensure adoption. Faculty and staff may resist AI integration due to concerns about job displacement, deskilling, or loss of professional autonomy. Effective change management requires transparent communication about AI's intended role as an assistant rather than a replacement.

Training programs must develop new competencies: prompt engineering, output verification, and AI-human collaboration workflows. Organizational culture must evolve to value efficiency gains while preserving the relational aspects of education that resist automation---mentorship, inspiration, and intellectual community.

\subsection{Future Directions}

Several research directions warrant exploration. Multi-agent architectures, where specialized agents coordinate to accomplish complex institutional tasks, may enable more sophisticated automation than single-agent systems. Personalized learning paths generated through AI analysis of student performance data could improve outcomes while reducing advising burdens. Predictive analytics may enable early intervention for at-risk students before failure occurs.

The long-term vision is of universities that learn and adapt: institutions where AI continuously analyzes outcome data, identifies improvement opportunities, and proposes evidence-based interventions. Realizing this vision requires sustained investment in infrastructure, governance frameworks, and human-AI collaboration models.

\subsection*{Scope and Non-Claims}

\noindent\fbox{%
\parbox{0.97\textwidth}{%
\textbf{Scope and Non-Claims.}
To avoid misinterpretation, we explicitly clarify the scope of the proposed framework:

\begin{itemize}
    \item The framework does not assume uniform institutional readiness; its applicability depends on existing organizational, technical, and data maturity.
    \item Reported efficiency gains are contingent upon the availability of unified, high-quality institutional data and should be interpreted as indicative rather than universally guaranteed.
    \item Level~5 (full autonomy) is presented as an aspirational boundary condition rather than a near-term implementation target.
    \item Human oversight remains mandatory for high-stakes academic decisions, including grading, degree conferral, accreditation submissions, and student progression.
\end{itemize}
}}

\section{Conclusion}

This chapter has proposed a framework for understanding and implementing agentic AI in higher education, drawing an analogy to the levels of autonomous vehicles. The self-driving university is not a destination but a trajectory---a progressive automation of administrative processes that liberates human capacity for the irreducibly human work of teaching, mentorship, and research.

Three key enablers underpin this vision. First, unified data infrastructure must precede AI deployment; automation atop fragmented systems yields partial benefits at best. Second, agentic AI architectures---with perception, planning, action, and memory capabilities---provide the autonomous execution that simple chatbots cannot. Third, progressive autonomy with human oversight ensures accountability while capturing efficiency gains.

The path forward requires institutions to prioritize infrastructure consolidation before AI experimentation. The technologies described in this chapter exist today; what remains is the organizational will to build the foundations upon which they can operate effectively. Universities that undertake this work position themselves not merely to survive the AI transformation but to lead it.

\section*{Acknowledgments}
The author thanks MUST University (Mediterranean University of Science and Technology, Tunisia) for hosting the pilot deployment and providing access to institutional data. Special thanks to the faculty members who participated in efficiency assessments and provided feedback during the deployment. The technical development team is acknowledged for their contributions to the XEducation platform implementation.

\bibliographystyle{splncs04}
\bibliography{references}

\end{document}